# Enhanced Momentum with Momentum Transformers

Max Mason[1], Waasi A Jagirdar[2], David Huang[1], Rahul Murugan[2]
*1: Data Science Institute, Columbia University*
*2: Electrical Engineering, Columbia University*

## Abstract

*The primary objective of this research is to build a Momentum Transformer that is expected to outperform benchmark time-series momentum and mean-reversion trading strategies. We extend the ideas introduced in [6] to equities as the original paper primarily only builds upon futures and equity indices. Unlike conventional Long Short-Term Memory (LSTM) models, which operate sequentially and are optimized for processing local patterns, an attention mechanism equips our architecture with direct access to all prior time steps in the training window. This hybrid design, combining attention with an LSTM, enables the model to capture long-term dependencies, enhance performance in scenarios accounting for transaction costs, and seamlessly adapt to evolving market conditions, such as those witnessed during the Covid Pandemic. The main technical challenges we faced are some of the sins mentioned in the paper "Seven Sins of Quantitative Investing "[11] where we inadvertently faced initial challenges, such as the data not being truly Point-In-Time (PIT) due to issues like data leakage, look-ahead biases, and possibly even survivorship bias, all mentioned in [11]. Further technical challenges were faced in the computation necessary for this strategy. To address these, the time period trained and tested on was reduced to 7 years and only one changepoint lookback window of 21 was used. After rectifying all errors, our results show promise for a few years and are similar with the original paper[6] although our best performing model doesn't use changepoint detection. We average 4.14% returns which is similar to their results. Our Sharpe is lower at an average of 1.12 due to much higher volatility which may be due to stocks being inherently more volatile than futures and indices.*

## 1. Introduction

There are various occurrences in financial markets that contradict the efficient market hypothesis. Of these market anomalies, one of the most popular strategies is the momentum strategy. This strategy is based on the fact that stocks with large returns over a certain period tend to have higher average returns over the next period [1]. The Moving Average Convergence Divergence (MACD) indicator is a popular method used in momentum strategies that helps identify buy and sell signals based on the convergence and divergence of two different length moving averages [2]. However, traditional momentum strategies utilizing MACD often face challenges in adapting to rapidly changing market conditions and may fail to capture complex temporal dependencies in price data.

To address these limitations, researchers have utilized advancements in machine learning and artificial intelligence to apply a deep learning framework to this problem which can learn trends and size positions by directly optimizing the result with the Sharpe ratio [3, 4]. These *Deep Momentum Networks* typically comprise of a Long Short-Term Memory (LSTM) model and have been shown to outperform classical momentum strategies by Wood et al. [4]. However, although the LSTM has been shown to perform well in short-term patterns, it tends to struggle with long-term patterns and responding to significant changes which can be seen in its poor performance in times of non-stationarity / momentum change points such as the Covid Pandemic [4,5]. One implementation used to improve this was a changepoint detection module to help the model identify regime change. This module is still limited in its ability to only use short-term information and not learn long-term patterns from potentially similar prior regimes [4].

Recent research by Wood et al. used a Transformer architecture, utilizing its multi-headed attention mechanism, to allow the model to learn both short-term and long-term dependencies [6]. When applied to time series data, attention mechanisms provide a learnable weight to measure the importance of prior time steps [7]. This allows the model to respond to specific events and learn regime-specific dynamics. This research found this methodology to outperform typical momentum strategies[6].

Prior research using Deep Momentum Networks has used commodities, indices, fixed income, and fx assets [3, 4, 6]. This paper aims to apply this Deep Momentum Network methodology to equities in recent years to see if the prior results still hold with assets where their data is more readily available and there are fewer barriers to entry to trade such assets.

## 2. Summary of the Original Paper
## 2.1 Methodology of the Original Paper

This paper is based on *Trading with the Momentum Transformer: An Intelligent and Interpretable Architecture* by Wood et. al [6] in which a Decoder-Only Temporal Fusion Transformer (TFT) is used to produce the positions at each time step.

This framework builds on the LSTM Deep Momentum Network architecture by using an LSTM encoding layer from prior research which is fed through a Variable Selection Network which filters out variables with low signal rates. This is then fed to a Gated Linear Uni(GLU)t which is another learned parameter through training that suppresses components to reduce complexity [8] and is always followed by an add & norm with skip connections that keep gradients from vanishing [9].

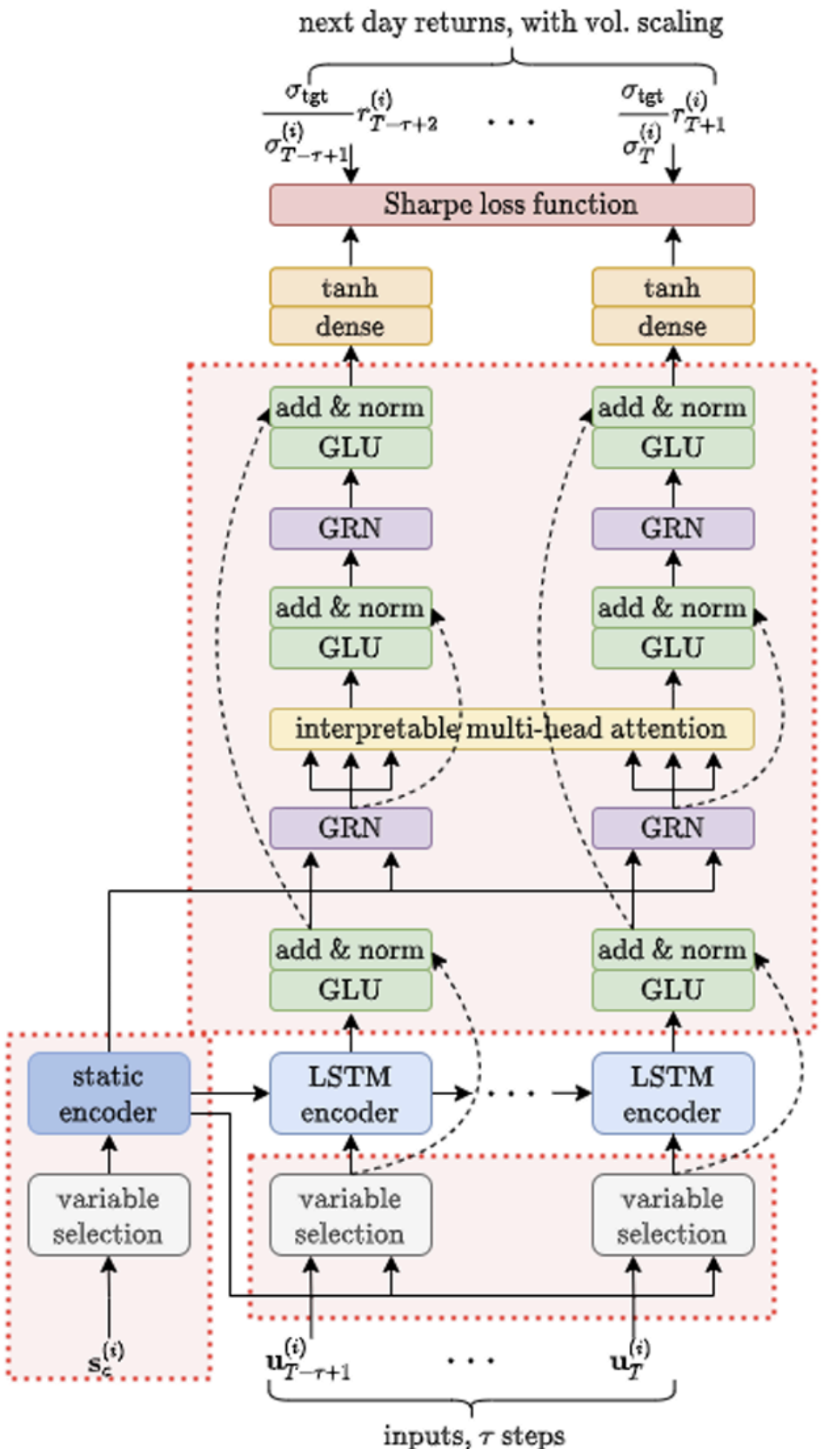

Image 1: TFT architecture [6]

The next layer is a Gated Residual Network(GRN) which passes the input through a feedforward neural network and applies a non-linear transformation. This output is then passed through a GLU which is then added and normed with the initial input data, allowing this block of the model to learn whether to apply non-linearity or not through this gating mechanism [10].

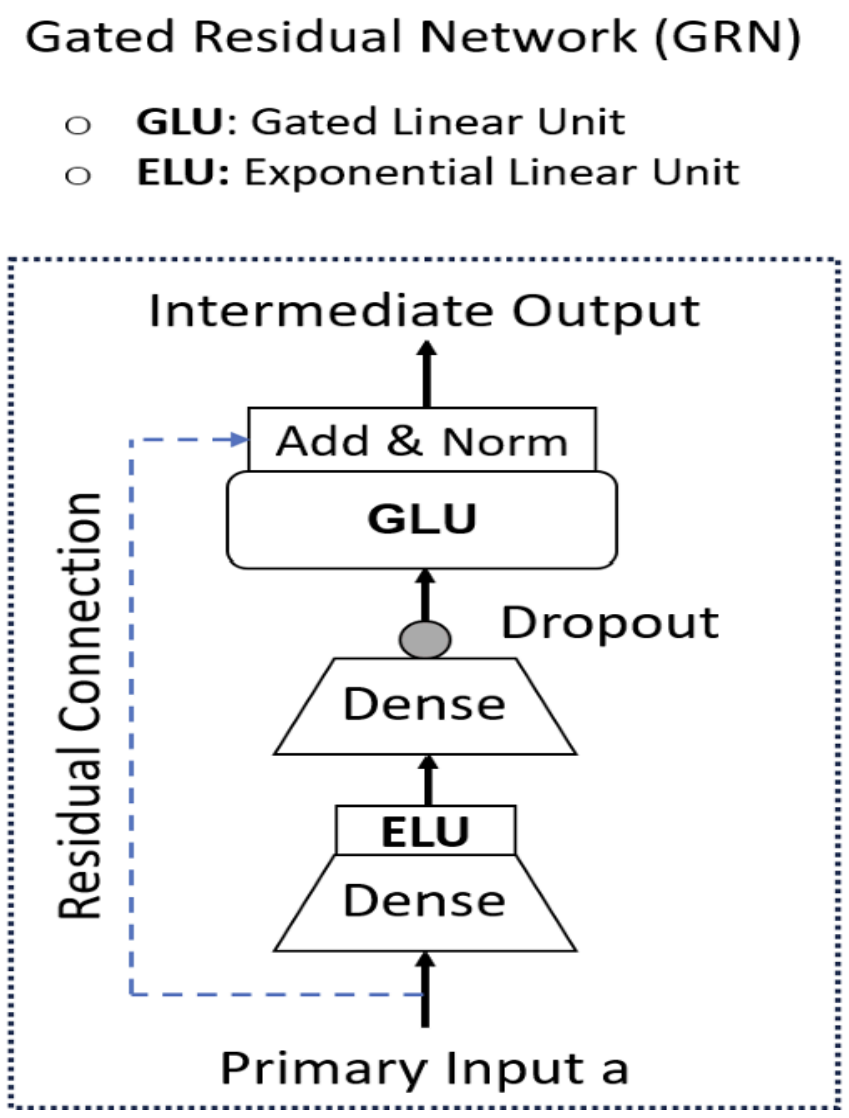

Image 2: Gated Residual Network Architecture [10]

This is then fed to a multi-head attention block that allows the model to capture multiple relationships among the input sequence. The GLU and GRN blocks are then repeated and the final output is fed to a dense feed-forward network with a tanh activation function. The output of this is the position for a given asset at a given timestep which is then used to calculate the Sharpe ratio as the loss function. The model is trained to minimize the negative Sharpe ratio [6].

The inputs of the model $u_T$ are the inputs at a given iteration of the model where $T$ represents the time steps included. In this paper, the TFT uses a window of size 252 to allow the multi-headed attention block to learn importances and relationships over a trading year period. They also include a static encoder in which the input variable $s_c$ is the asset class so their model can learn behavioral differences in the different assets they trade. Wood et. al focus primarily on trading futures due to their low covariance compared to equities. The portfolio in this original paper was comprised of 25 commodities, 11 indices, 5 fixed income, and 9 foreign exchange assets.

## 2.2 Key Results of the Original Paper

They were able to improve upon prior research with the TFT model producing a Sharpe ratio of 2.62 from 1995 to 2020. It was also shown that this model performed well in times of market turmoil as this new model had a Sharpe of 2.47 during the Covid Pandemic while prior Deep Momentum Networks utilizing the LSTM suffered with a Sharpe of -1.50 during the same period.

## 3. Methodology

The goal of this paper is to test this strategy and implementation with only equities. With this in mind, we tried to reduce covariance as much as possible by equally diversifying our assets across sectors. Our model's inputs are all derived directly from price data, specifically the price at close. The model is trained to make the portfolio allocation, predicting the next day's close based on variables including the current day's close and past close prices. Since we are using close prices this allows us to make our portfolio rebalances at or near close, after hours, or when the market opens allowing for tradability of the strategy.

Furthermore, in recent years such as 2020, 2022, and 2023 momentum strategies have struggled to make risk-adjusted returns due to high market volatility. We plan to expand the time steps included in each iteration of the model from 252 to 378 to allow the attention mechanism to learn further temporal dynamics over one and half years of trading rather than one to see if this can improve performance in periods where momentum strategies struggle.

Another method variation we test during this period is an increased number of attention heads. The original paper utilizes 4 attention heads but by increasing this we allow the model to capture more aspects of the data simultaneously, allowing the model to learn more diverse patterns and generalize better.

### 3.1. Technical Challenges

Initially, we faced challenges with data leakage that caused our results to exceed possible Sharpe values. The issue arose from certain companies in our portfolio that had multiple versions of their stock. This caused the initial backward shift of one trading day for our model's target variable in training for these companies to not work properly as they had multiple instances on each trading day, giving the model the ability to see future values for these companies. Once found and fixed, our model outputs were proper, and the data point-in-time.

Due to computational limitations, we could only use changepoint detection for a lookback window length of 21. The original paper utilized changepoint detection of both windows of size 21 and 126 to get short-term and long-term regime shifts. This, however, wasn't feasible as the lookback window of 21 already took on average 30 minutes per company and required above 12 GB of RAM for each. When tested, the lookback window of size 126 typically required double this. This also caused us to reduce our sample size from 10 years to 7, reducing the amount of training data. To combat overfitting we increased the validation set size from 10% to 20%. Lastly, when testing the increase in time steps included in each iteration we would have ideally used a larger increase to two years but this would cause our run to crash as we run out of GPU space.

## 4. Data

For our project, we use data filtered to include the **top 5 companies by market capitalization** within specific ranges of **SIC (Standard Industrial Classification) codes**, which correspond to various broad industry sectors. The SIC code ranges for these industry sectors are defined as follows:

SIC 0100-0199 Agriculture, Forestry and Fishing
SIC 1000-1499 Mining
SIC 1500 -1799 Construction
SIC 2000-3999 Manufacturing
SIC 4000-4999 Transportation, Communications, Electric, Gas, and Sanitary Service
SIC 5000 -5199 Wholesale Trade
SIC 5200-5999 Retail Trade
SIC 6000-6799 Finance, Insurance and Real Estate
SIC 7000-8999 Services.

The data sources used include CRSP and Compustat and the primary identifier used is the 'cusip' column. The identifier allows us to join tables through SQL to extract relevant information, which in our case are the top 5 companies sectorally by market capitalization. Our final dataset contains time series data for the filtered companies including relevant columns such as 'cusip', 'ticker', 'date', 'market_cap', 'returns', and 'sic code'.
Based on the unique tickers identified through this dataframe, we run the Change-point detection (CPD) script which does the following,

1. Fetch daily return data for the stock corresponding to the specified ticker from the **CRSP (Center for Research in Security Prices)** database within a specific date range.
2. Organize and format the data for further analysis by extracting relevant columns and renaming them for clarity(daily returns).
3. Analyze the stock's return time series for structural changes (e.g., regime shifts, anomalies) using a specified **lookback window, for example**, 126 days.
4. **The final columns obtained after this step include,**
   1. date (the actual date associated with the detected change points or time-series data.
   2. t (The time index or sequential number corresponding to the date in the time series.)
   3. cp_location (The exact location (index) in the time series where a **change point** is detected. i.e,

indicates the time step where there is a significant shift in the statistical properties of the data)

4. cp_location_norm (The **normalized location** of the change point within the analyzed time window.)

5. cp_score (The **change point score**, which quantifies the strength or significance of the detected change point.)

# 5. Results

## 5.1 Project Results

Each model version was trained on an initial in-sample training data and validation set from the first three years of data. It is then tested on out of sample the next year using a sliding window to make predictions each day. The out of sample year is then added to the training data, a new model is trained, and the new model is again tested on the next out of sample year. This is an expanding window approach where the initial training data is from 2017-2019, then 2017-2020, and so on until 2022 where our final result is tested out of sample in 2023. With this we get results of our different tested methods from 2020 to 2023, giving us insight into how each performs per year in multiple market conditions as well as a four year average of our results. Over this time period the classical momentum strategy we sought to enhance struggled, averaging annual returns of -1.07% and an average Sharpe of -0.18. Interestingly enough, it was found that long only performed quite well in this period, with an average annual return of 4.04% but an average Sharpe of 0.57. The overall best performing model was the vanilla TFT model from the original paper [6] which produced average annual returns of 4.14% and an average annual Sharpe of 1.12. The variations we added with the expanded window and increased attention heads negatively impacted the non-CPD TFT. The larger training window also negatively impacted the model with CPD so it's believed that this caused some overfitting issues. However, it appears that increasing the number of attention heads did improve the performance of the TFT-CPD showing promise that more attention heads may be able to help the model discern changes in noisy data. These results and the rest from all our tests can be seen in table 1 below by year and the overall average.

Table 1: Strategy Performance

| | Returns | Vol. | Sharpe | Downside Risk | Sortino | Maximum Drawdown | Calmar | % Positive Returns | Profit/Loss Ratio |
|---|---|---|---|---|---|---|---|---|---|
| **2020** | | | | | | | | | |
| Long-Only | 0.47% | 11.31% | 0.09 | 8.85% | 0.13 | 16.00% | 0.03 | 56.12% | 0.796 |
| Momentum | -9.35% | 5.85% | -1.65 | 5.03% | -1.92 | 11.31% | -0.83 | 53.75% | 0.626 |
| LSTM-CPD | 0.23% | **3.33%** | 0.09 | **2.50%** | 0.11 | **4.90%** | 0.05 | 52.17% | **0.934** |
| $TFT_{252, 4}$ | **8.42%** | 7.42% | 1.13 | 5.36% | 1.56 | 8.51% | 0.99 | 58.10% | 0.933 |
| $TFT_{378, 4}$ | 6.36% | 8.47% | 0.77 | 6.44% | 1.01 | 10.39% | 0.61 | 57.31% | 0.890 |
| $TFT_{252, 6}$ | 7.88% | 8.10% | 0.98 | 5.97% | 1.33 | 10.16% | 0.78 | **59.29%** | 0.848 |
| $TFT\text{-}CPD_{252, 4}$ | 4.68% | 7.19% | 0.67 | 5.35% | 0.90 | 8.53% | 0.55 | 56.52% | 0.890 |
| $TFT\text{-}CPD_{378, 4}$ | 3.37% | 5.32% | 0.65 | 3.92% | 0.88 | 5.15% | 0.65 | 58.10% | 0.823 |
| $TFT\text{-}CPD_{252, 6}$ | 5.57% | 4.35% | **1.27** | 3.21% | **1.72** | 5.52% | **1.01** | **59.29%** | 0.877 |
| **2021** | | | | | | | | | |
| Long-Only | **10.93%** | 7.82% | 1.36 | 5.51% | 1.94 | 4.20% | 2.60 | 56.35% | 0.968 |
| Momentum | 6.59% | 5.82% | 1.13 | 4.19% | 1.56 | 3.45% | 1.91 | **57.54%** | 0.886 |
| LSTM-CPD | 7.46% | 5.14% | 1.43 | 3.54% | 2.07 | 2.80% | 2.67 | 57.14% | 0.949 |
| $TFT_{252, 4}$ | 3.11% | **1.58%** | **1.95** | **1.05%** | **2.93** | **1.11%** | 2.79 | 56.74% | 1.060 |
| $TFT_{378, 4}$ | 3.54% | 3.21% | 1.10 | 2.22% | 1.59 | 1.92% | 1.84 | 52.78% | 1.082 |
| $TFT_{252, 6}$ | 6.53% | 5.18% | 1.25 | 3.39% | 1.91 | 2.55% | 2.56 | 52.78% | 1.109 |
| $TFT\text{-}CPD_{252, 4}$ | 6.50% | 3.76% | 1.69 | 2.60% | 2.45 | 2.36% | 2.76 | **57.54%** | 0.978 |
| $TFT\text{-}CPD_{378, 4}$ | 9.29% | 5.58% | 1.62 | 3.63% | 2.49 | 2.77% | **3.36** | 53.17% | **1.157** |
| $TFT\text{-}CPD_{252, 6}$ | 6.54% | 4.35% | 1.48 | 2.89% | 2.22 | 2.04% | 3.20 | 55.56% | 1.030 |
| **2022** | | | | | | | | | |
| Long-Only | -5.43% | 9.62% | -0.53 | 6.97% | -0.73 | 11.03% | -0.49 | 49.00% | 0.954 |
| Momentum | -2.02% | 5.30% | -0.36 | 4.12% | -0.46 | 4.14% | -0.49 | **50.99%** | 0.907 |
| LSTM-CPD | -0.83% | 2.98% | -0.26 | 2.09% | -0.38 | 2.96% | -0.28 | 48.21% | 1.025 |
| $TFT_{252, 4}$ | 0.53% | **1.80%** | 0.31 | **1.13%** | **0.49** | **1.66%** | 0.32 | 47.01% | 1.189 |
| $TFT_{378, 4}$ | -1.93% | 2.42% | -0.79 | 1.74% | -1.11 | 2.70% | -0.72 | 45.42% | 1.052 |
| $TFT_{252, 6}$ | -1.38% | 3.91% | -0.34 | 2.56% | -0.51 | 4.19% | -0.33 | 46.22% | 1.097 |
| $TFT\text{-}CPD_{252, 4}$ | **0.71%** | 4.87% | 0.17 | 3.48% | 0.26 | 3.48% | 0.21 | 46.22% | 1.199 |
| $TFT\text{-}CPD_{378, 4}$ | -1.85% | 3.18% | -0.57 | 2.17% | -0.84 | 3.61% | -0.51 | 45.42% | 1.091 |
| $TFT\text{-}CPD_{252, 6}$ | 1.15% | 3.72% | **0.32** | 2.48% | **0.49** | 2.99% | **0.38** | 46.22% | **1.23** |
| **2023** | | | | | | | | | |
| Long-Only | **10.17%** | 7.43% | **1.34** | 4.89% | **2.04** | 7.63% | **1.33** | **55.82%** | 0.978 |
| Momentum | 0.50% | 3.36% | 0.17 | 2.41% | 0.23 | 3.50% | 0.14 | 49.40% | 1.043 |
| LSTM-CPD | 2.87% | 4.03% | 0.72 | 2.60% | 1.12 | 5.08% | 0.56 | 51.00% | **1.087** |
| $TFT_{252, 4}$ | 4.49% | 4.11% | 1.09 | 2.85% | 1.57 | 4.34% | 1.03 | 55.42% | 0.963 |
| $TFT_{378, 4}$ | 2.16% | 2.75% | 0.79 | 1.84% | 1.19 | **1.63%** | **1.33** | 51.41% | 1.083 |
| $TFT_{252, 6}$ | 2.36% | 4.92% | 0.50 | 3.41% | 0.72 | 4.63% | 0.51 | 52.21% | 0.996 |
| $TFT\text{-}CPD_{252, 4}$ | 2.60% | 2.70% | 0.97 | 1.83% | 1.42 | 2.38% | 1.09 | 52.21% | 1.073 |
| $TFT\text{-}CPD_{378, 4}$ | 2.10% | **2.59%** | 0.82 | **1.72%** | 1.23 | 2.28% | 0.92 | 52.21% | 1.058 |
| $TFT\text{-}CPD_{252, 6}$ | 2.60% | 4.84% | 0.55 | 3.28% | 0.81 | 4.41% | 0.59 | 52.21% | 1.002 |
| **Average** | | | | | | | | | |
| Long-Only | 4.04% | 9.05% | 0.57 | 6.56% | 0.85 | 9.72% | 0.87 | **54.32%** | 0.924 |
| Momentum | -1.07% | 5.08% | -0.18 | 3.94% | -0.15 | 5.60% | 0.18 | 52.92% | 0.866 |
| LSTM-CPD | 2.43% | 3.87% | 0.49 | 2.68% | 0.73 | 3.94% | 0.75 | 52.13% | 0.999 |
| $TFT_{252, 4}$ | **4.14%** | **3.72%** | **1.12** | **2.59%** | **1.64** | **3.90%** | 1.28 | **54.32%** | **1.036** |
| $TFT_{378, 4}$ | 2.53% | 4.21% | 0.47 | 3.06% | 0.67 | 4.16% | 0.77 | 51.73% | 1.027 |
| $TFT_{252, 6}$ | 3.85% | 5.53% | 0.60 | 3.83% | 0.86 | 5.38% | 0.88 | 52.62% | 1.013 |
| $TFT\text{-}CPD_{252, 4}$ | 3.62% | 4.62% | 0.88 | 3.23% | 1.26 | 4.19% | 1.15 | 53.12% | 1.035 |
| $TFT\text{-}CPD_{378, 4}$ | 3.23% | 4.17% | 0.63 | 2.86% | 0.94 | 3.45% | 1.10 | 52.23% | 1.032 |
| $TFT\text{-}CPD_{252, 6}$ | 3.96% | 4.32% | 0.91 | 2.97% | 1.31 | 3.74% | **1.29** | 53.32% | 1.034 |

*The subscript numbers for the transformer models are {training window timesteps, number of attention heads

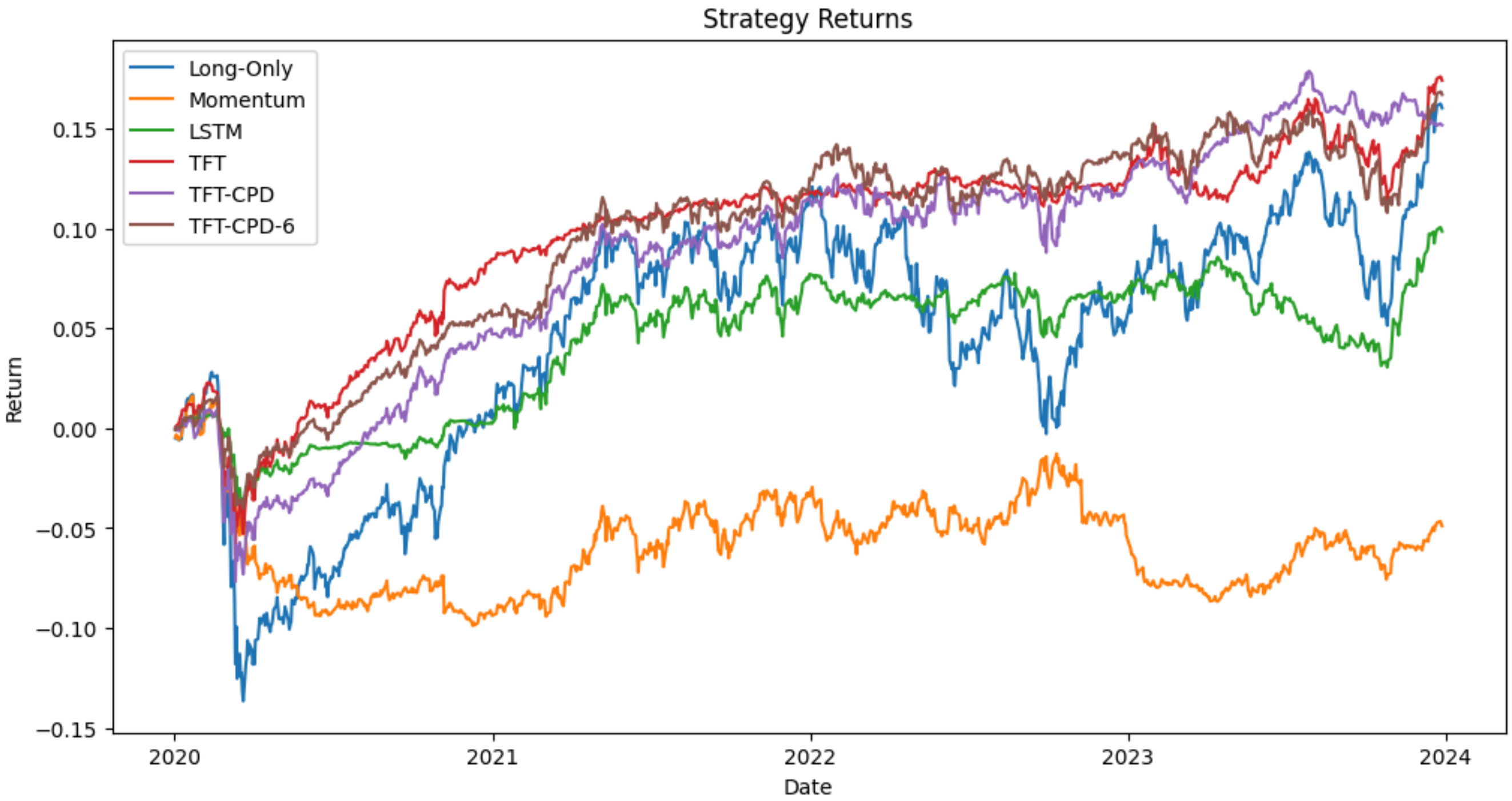

Image 3: Cumulative Returns of top performing strategies and baseline

Over time, the top performing TFT and TFT with 6 attention heads and changepoint detection both barely outperform the long-only for this whole testing period. However, we can clearly see that they are less impacted by market changes as when there are large drawdowns in the long-only our top performing models are more stable and not as impacted by these market effects. This makes sense as our models are optimized to increase Sharpe which takes into account the volatility of the return so we would expect our models to be less volatile.

It's hard to fully compare these results with the original paper because we're looking at two different time periods, but overall their long-term test showed average annual returns of 4.01% which we also achieved. We, however, had much higher volatility with their TFT models achieving a volatility of 1.51% even in a highly volatile market during the Covid Pandemic [6].

## 6. Discussion & Future Work

A large cause for our reduced Sharpe compared to their results stems from the different portfolio we tried trading which is highly more volatile than theirs. This can be seen when we compare long-only strategy volatilities. During the Covid Pandemic they're long-only strategy had an annual volatility of 6.73% [6]. Our portfolio on the other hand, never went below an annual volatility of 7.43% and over the four years averaged an annual volatility of 9.05%. Strategies such as increasing the time window input for the model or increasing the number of attention heads were used to try and counteract this but provided minimal or worse results. Further volatility mitigation strategies were tested including doubling the number of stocks in the portfolio from each sector to ten but this caused a largely negative result. This is likely due to the fact that equities tend to exhibit high covariance. We tried to mitigate this by diversifying sectorally but then each of the five companies from that sector exhibit high covariance with each other. This idea, however, falls short after looking into the annual covariances of returns for our portfolio which all appear reasonably low.

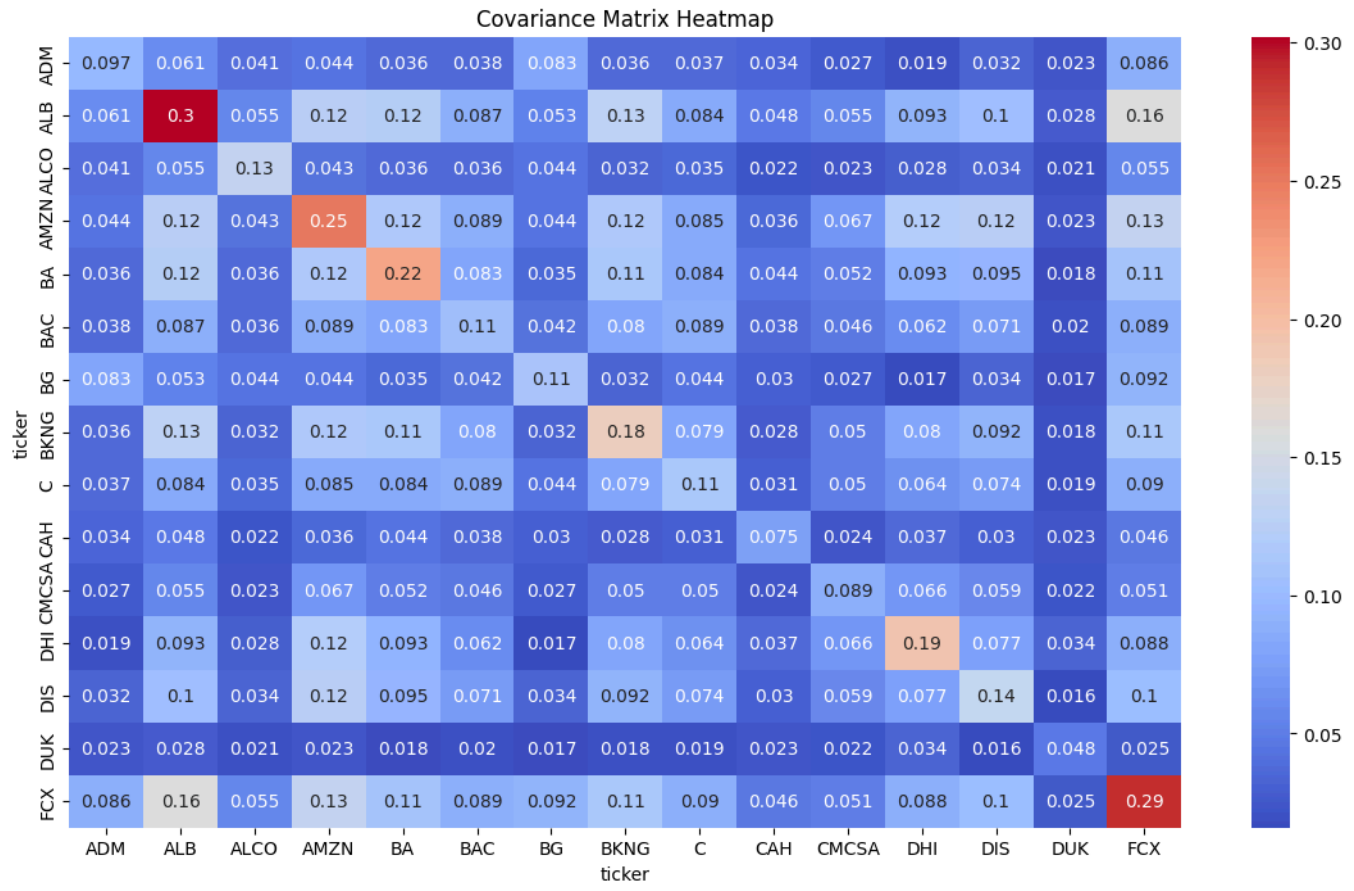

Image 4: Covariance Matrix Heatmap for 15 companies in 2022

Another potential issue we noticed was the CPD was almost always producing a high confidence change. It's believed that this was due to how volatile equities can be and likely why the model did better without the CPD.This lead to the idea that maybe more attention heads could discern something from the noise that was the Changepoint features but it only narrowly improved the model and still didn't beat the model without it. We also didn't have the long-term CPD which would have helped in long-term trend changes which we can actually see a few occur during this time.

Future work should look to include this long-term CPD with a larger number of attention heads for the model to better determine temporal relationships among returns, MACD, and CPD features. Furthermore, when trading with equities it may be prudent to include other important features commonly used in factor models such as variables for the size, value, and market factors to help the model learn and reduce market exposure, potentially decreasing volatility.

## 7. Summary

This paper looked to improve upon common momentum strategies by trading equities using a Momentum Transformer used in prior research. This model takes in classical momentum strategy features and learns a best portfolio for the next day. The goal was to use the methodology to improve the basic momentum strategy that used MACD indicators. This was achieved as over a four year period our model outperformed the basic momentum strategy by an average annual return of 5.21% and an increase in the Sharpe ratio of 1.3. This, however, doesn't compare to prior research results which was able to achieve a Sharpe of 2.00 versus our best of 1.12. Furthermore, our average annual return was only slightly better than the long-only return in this same period, putting into question the added overall performance of our model. It is believed that these observed low returns may reflect unique challenges during the testing period, such as high volatility stemming from the Covid Pandemic, rising interest rates, and rising inflation. These results also lead us to conclude that, although we achieved an annual return of 4.14% and Sharpe of 1.12, momentum strategies are likely better geared towards less volatile assets such as futures and indices.

## 6. Acknowledgement

We would like to thank Professor Naftali Cohen for his class and feedback throughout this project. We would also like to thank the Oxford-Man Institute of Quantitative Finance for their research and code on which our project is based upon.

## 7. References

github:
https://github.com/MaxMason42/DDMIF_Final_Project

### 8.2 Support Material

The github for the original paper which this paper was based on:
https://github.com/kieranjwood/trading-momentum-transformer